\documentclass[conference]{IEEEtran}
\usepackage{graphicx}
\usepackage{cite}
\usepackage{amsmath}
\usepackage{amsfonts}
\usepackage{amssymb}
\usepackage{subfigure}
\usepackage{psfrag}
\usepackage{epstopdf}

\newcommand{\bm}{\mathbf}
\newcommand{\be}{\begin{equation}}
\newcommand{\ee}{\end{equation}}
\newcommand{\bea}{\begin{eqnarray}}
\newcommand{\eea}{\end{eqnarray}}

\newcommand{\bA}{{\bm A}}

\newcommand{\bR}{{\bm R}}
\newcommand{\bW}{{\bm W}}

\newcommand{\bS}{{\bf S}}
\newcommand{\bH}{{\bf H}}

\newcommand{\bh}{{\bf h}}
\newcommand{\bv}{{\bf v}}
\newcommand{\bw}{{\bf w}}

\newcommand{\bs}{{\bf s}}
\newcommand{\bx}{{\bf x}}
\newcommand{\br}{{\bm r}}

\newcommand{\I}{{\bm I }}

\newcommand{\bPhi}{\mbox{\boldmath{$\Phi$}}}

\title{Frequency Spreading Equalization in \\Multicarrier Massive MIMO }

\author{\normalsize Amir Aminjavaheri$^\dagger$, Arman Farhang$^*$, Nicola Marchetti$^*$, Linda E. Doyle$^*$ and Behrouz Farhang-Boroujeny$^\dagger$  
\\$^*$CTVR / The Telecommunications Research Centre, Trinity College Dublin, Ireland, \\
$^\dagger$ECE Department, University of Utah, USA. \\
Email: \{farhanga, marchetn, ledoyle\}@tcd.ie, \{aminjav, farhang\}@ece.utah.edu \vspace{-0.2in}}

\begin{document}

\maketitle

\begin{abstract}
Application of filter bank multicarrier (FBMC) as an effective method for signaling over massive MIMO channels has been recently proposed. This paper further expands the application of FBMC to massive MIMO by applying frequency spreading equalization (FSE) to these channels. FSE allows us to achieve a more accurate equalization. Hence, higher number of bits per symbol can be transmitted and the bandwidth of each subcarrier can be widened. Widening the bandwidth of each subcarrier leads to (i) higher bandwidth efficiency; (ii) lower complexity; (iii) lower sensitivity to carrier frequency offset (CFO); (iv) reduced peak-to-average power ratio (PAPR); and (iv) reduced latency. All these appealing advantages have a direct impact on the digital as well as analog circuitry that is needed for the system implementation. In this paper, we develop the mathematical formulation of the minimum mean square error (MMSE) FSE for massive MIMO systems. This analysis guides us to decide on the number of subcarriers that will be sufficient for practical channel models.

 \vspace{-0.1in}

\end{abstract}

\section{Introduction}\label{sec:intro}
Based on the recent discussions on the $5^{\rm th}$ generation of wireless communication networks (5G), massive MIMO, as a candidate, can play a pivotal role in these networks, \cite{5G}. Additionally, as a fundamental building block in the physical layer of 5G networks, the need for signaling techniques that are compatible with the emerging applications such as the \textit{machine to machine} (M2M) communications and \textit{Internet of Things} (IoT) is greatly emphasized \cite{5GWaveforms2014}. Filter bank multicarrier (FBMC), a method that was initially proposed about 50 years ago, \cite{Chang66,Saltzberg67}, is recently being considered as a candidate for 5G systems, \cite{5G}. 

Massive MIMO is a code division multiple access (CDMA) like system that may be used to increase the capacity of multiuser networks, \cite{MMIMOsurver2013}. As it is shown in \cite{Marzetta2010}, by increasing the number of antennas at the base station (BS), the effects of noise and multiuser interference (MUI) start to vanish until they will be completely removed as the number of BS antennas tends to infinity. In the recent works \cite{FBMCMassive2014} and \cite{AFISWCS14}, FBMC was proven to be a good match for massive MIMO and vice versa as they can both bring a great number of appealing properties into the picture of 5G systems. From network capacity viewpoint, this combination is of a paramount importance as the same spectrum is not only being reused by all the users (an advantage of massive MIMO) but it is being used in a more bandwidth efficient way.

A few recent publications have introduced an alternative to the traditional polyphase-based implementation of the FBMC systems that is termed as frequency spreading FBMC (FS-FBMC), \cite{FBMCPrimer,FS-FBMCISCCSP2012,FS-FBMC2012,FS-FBMC2014, Amin1506:Frequency}. This technique offers a number of advantages that put FBMC in a strong position as a candidate for the future wireless standards. FS-FBMC was explained for the first time in a document that came out of the PHYDYAS project, \cite{FBMCPrimer}. Further details of FS-FBMC were presented, by the same author, in \cite{FS-FBMCISCCSP2012}. The sensitivity of FS-FBMC to the timing offset is studied in \cite{FS-FBMC2012} where the robustness of the FS-FBMC frequency domain equalizer to the timing errors compared with its counterpart with a polyphase-based receiver structure was reported.
 Berg et al., \cite{FS-FBMC2014}, have explored the utilization of FS-FBMC to access TV white spaces. All these works have used the filter designed by Martin \cite{Martin98} and Mirabbasi and Martin \cite{Mirabbasi2003} for the implementation of FS-FBMC system.

As an extension to our previous work, \cite{FBMCMassive2014}, in this paper, we introduce the frequency spreading equalization (FSE) concept to the realm of FBMC-based massive MIMO systems. Thanks to the better equalization capability of FS-FBMC systems, we will show that FSE brings a great amount of performance enhancement compared with the results of \cite{FBMCMassive2014}. This allows the usage of even wider subcarriers than what was suggested in \cite{FBMCMassive2014} which in turn further reduces the latency due to the synthesis and analysis filter banks. Furthermore, this unlocks utilization of higher order modulation schemes in FBMC-based massive MIMO systems. Using a smaller number of subcarriers leads to a lower complexity, lower sensitivity to CFO, reduced PAPR and reduced latency. These advantages simplify the digital and analog circuitry that is needed for the system implementation.

Widening the bandwidth of each subcarrier is equivalent to shortening the length of the prototype filter. This, in turn, has the impact of increasing the bandwidth efficiency of FBMC, as the ramp-up and ramp-down of the FBMC signal will be shortened. A factor that limits utilization of FBMC systems in some applications like bursty M2M communications is the long ramp-up and ramp-down of their signals. To tackle this problem, some researchers have introduced the concept of FBMC systems with circular pulse-shaping, \cite{GFDM,Arman_LowComplexity,COQAM,CB-FMT}, where the prototype filter transients can be removed through the so called \textit{tail biting property}. Even though these systems can remove the prototype filter transients from their signals, they need to add a cyclic extension to form their signal. Additionally, they suffer from higher out of band emissions than the FBMC systems with linear pulse-shaping, \cite{Fettweis_LowComplexityTx,COQAM}. The approach in this paper is to reduce the ramp-up and ramp-down of the FBMC signal through widening the subcarriers while achieving a good equalization performance. This reduced transient duration of the linearly pulse-shaped FBMC signals is equivalent to the cyclic extensions being used to form the circularly pulse-shaped FBMC signals. Therefore, in terms of bandwidth efficiency, both systems seem to be about the same while linearly pulse-shaped FBMC systems keep the advantage of having a lower out of band emissions.

The rest of the paper is organized as follows. FBMC signal construction and the frequency spreading method are explained in Section~\ref{sec:FSCMT}. The proposed frequency spreading equalization for the massive MIMO systems is discussed in Section~\ref{sec:FSE} where we have also derived analytical signal to interference plus noise (SINR) expressions. 
Various aspects of our design are analyzed through computer simulations in Section~\ref{sec:NR}. Finally, the paper is concluded in Section~\ref{sec:Conclusion}.

\textit{Notations}: Scalars, vectors and matrices are represented in regular letters, boldface lower and upper case letters, respectively. $\mathbf{I}_N$ denotes an $N\times N$ identity matrix. The matrix or vector superscripts $(\cdot)^{\rm T}$ and $(\cdot)^{\rm H}$ indicate transpose and conjugate transpose, respectively. $\| \cdot \|$ and $|\cdot|$ represent Euclidean norm and absolute value, respectively. $\Re\{\cdot\}$, $\Im\{\cdot\}$, $\mathbb{E}[\cdot]$ and $(\cdot)^{-1}$ identify the real and imaginary parts of a complex number, expectation and inverse of a matrix, respectively. Linear convolution is denoted by $\star$, and the superscript $*$ represents the complex conjugate.

\section{CMT and its Frequency Spreading Implementation}\label{sec:FSCMT}

In cosine modulated multitone (CMT), which is a type of FBMC similar to the widely known OQAM-FBMC technique \cite{FarYuen2010}, real-valued data symbols are spread across time and frequency such that if the symbol time spacing, i.e., symbol period, is denoted by $T$, the symbol frequency spacing is $F=\frac{1}{2T}$. Moreover, subcarrier phases are toggled between $0$ and $\pi/2$ so that adjacent subcarriers have a quadrature phase difference. This \textit{phase-adjustment} is necessary for the CMT receiver to be able to recover the transmitted data symbols free of interference along time and frequency. The filter $p(t)$, which is used to pulse shape the data symbols, is a square-root Nyquist filter, i.e., $q(t) = p(t) \star p^*(-t)$ is a Nyquist filter, where the spacing between its zero-crossings is equal to $2T$.

In practice, the CMT signal is efficiently synthesized at the transmitter and analyzed at the receiver using the polyphase network (PPN) \cite{farhang2014filter, Farhang2011}. However, as it has been recently suggested by Bellanger \cite{FBMCPrimer}, the frequency spreading structure can be utilized to gain a number of advantages compared with the traditional PPN method. In particular, an interesting property of FS-FBMC structure is 
its high-performance equalization capability. This equalization, namely, 
FSE, can be utilized in the context of massive MIMO as suggested in Section \ref{sec:FSE}. In the rest of this section, and for our later reference, we review the frequency spreading structure when applied to CMT, which we hereafter call \textit{FS-CMT}. 

The basic idea of the frequency spreading method is that a discrete-time square-root Nyquist filter, $p[n]$, of size $N=KL$, can be synthesized in the frequency domain using a set of $2K-1$ distinct tones, as
\bea
p(n) = \sum_{k = -K+1}^{K-1} c_k e^{j\frac{2\pi k n}{N}} .
\eea
Here, $c_k$ is a real-valued coefficient that indicates the frequency response of the $p(n)$ at the frequency $\omega_k = \frac{2\pi k}{N}$, $K$ is the overlapping factor, i.e., it signifies the number of symbols that overlap in the time domain and $L$ is the sample spacing of the zero-crossings of the Nyquist filter $q(n)=p(n) \star p^*(-n)$.

When the above idea is extended to the multicarrier case, the symbols at each subcarrier should be pulse-shaped in the frequency domain using the set of $c_k$'s. An inverse discrete Fourier transform (IDFT) block of size $N$ can then be used to calculate the time domain samples from their frequency domain counterparts. Casting the above discussion into the mathematical form, the FS-CMT symbol can be obtained as 
\bea
\bx(n) = \mathbf{F}_N^{\rm H} \bA \bPhi \bs(n) .
\eea
Here, $\bs(n) = \begin{bmatrix} s_0(n) & s_1(n) & \cdots & s_{L-1}(n) \end{bmatrix}^{\rm T}$, where $s_k(n)$ indicates the data symbol at the $k^{\rm th}$ subcarrier and $n^{\rm th}$ symbol time, $\bPhi$ indicates the phase-adjustment matrix which is an $L\times L$ diagonal matrix with the diagonal elements $\{1,e^{j\frac{\pi}{2}},\dots\,e^{j\frac{\pi}{2}(L-1)}\}$, $\bA$ is the spreading matrix of size $N\times L$ whose $k^{\rm th}$ column contains the coefficients $c_\ell$'s that are centered at the
 center frequency of the $k^{\rm th}$ subcarrier, $\mathbf{F}_N^{\rm H}$ indicates the IDFT matrix and $\bx(n)$ is an $N \times 1$ vector indicating the $n^{\rm th}$ FS-CMT symbol in time domain. 

\begin{figure}[!t]
\centering
\includegraphics[scale=0.9]{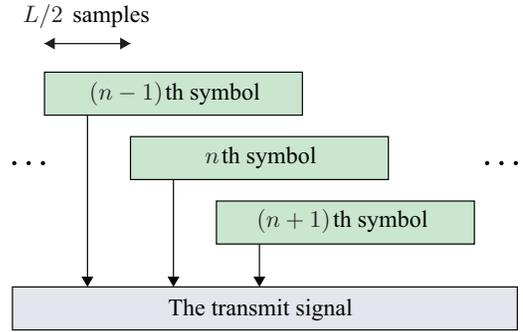}
\caption{The overlap and add operation.}
\label{fig:OA} \vspace{-0.2in}
\end{figure}

Noting that the size of each FS-CMT symbol is $N$ and the symbol spacing is equal to $L/2$, the transmit signal can be constructed by performing the overlap-and-add operations on the symbol vectors as illustrated in Fig.~\ref{fig:OA}. Assuming an ideal channel and having a perfect synchronization between transmitter and receiver, the $N$ samples that correspond to the $n^{\rm th}$ symbol are collected in the vector $\hat{\br}(n)$, at the receiver side, and the data symbols are recovered according to
\be
\hat{\bs}(n) = \Re \left\{ \bPhi^{-1} \bA^{\rm T} \mathbf{F}_N  \hat{\br}(n) \right\} .
\ee
Since we assumed an ideal channel, equalization was not considered here. In the presence of the channel, the equalization can be embedded after obtaining the DFT output samples, \cite{FS-FBMCISCCSP2012}. In the next section, we extend the equalization scheme that was originally proposed for single antenna systems in \cite{FS-FBMCISCCSP2012} to the context of massive MIMO.

\section{MMSE-FSE in Massive MIMO}\label{sec:FSE}

{\begin{figure*}[!t]
{
\setcounter{equation}{7}
{\be\label{eqn:PS}
P_{\rm s}^{(\ell,i)}=\sum^{K-1}_{k=-K+1}{c_k^2}|\bw_{i+k,{\Re}}^{\rm T}(\ell)\bh_{i+k,{\Re}}(\ell)+\bw_{i+k,{\Im}}^{\rm T}(\ell)\bh_{i+k,{\Im}}(\ell)|^2,
\ee}
\bea\label{eqn:PI}
P_{\rm I}^{(\ell,i)}&=&\sum^{K-1}_{k=-K+1}{\sum\limits_{\substack{m=0 \\ m \neq \ell}}^{M-1}}{c_k^2}|\bw_{i+k,{\Re}}^{\rm T}(\ell)\bh_{i+k,{\Re}}(m)+\bw_{i+k,{\Im}}^{\rm T}(\ell)\bh_{i+k,{\Im}}(m)|^2 \nonumber \\
 &&+ \sum^{K-1}_{k=-K+1}\sum^{M-1}_{m=0}{c_k^2}|\bw_{i+k,{\Re}}^{\rm T}(\ell)\bh_{i+k,{\Im}}(m)-\bw_{i+k,{\Im}}^{\rm T}(\ell)\bh_{i+k,{\Re}}(m)|^2+\sum^{K-1}_{k=-K+1}{c_k^2}\sigma_v^2\|\bw_{i+k}(\ell)\|^2.
\eea}
\setcounter{equation}{3}
\hrulefill
\end{figure*}
}

An interesting property of FBMC in massive MIMO channels that is called \textit{self-equalization} was recently reported in \cite{FBMCMassive2014}. Due to that property, a channel flattening effect takes place as a result of linear combining of the channel gains in different receiver antennas during the equalization procedure. As a result, wider subcarriers in frequency can be utilized which brings a number of benefits to FBMC systems in massive MIMO application. The benefits of FBMC-based massive MIMO systems are well articulated in \cite{FBMCMassive2014} and \cite{AFISWCS14}. As it is shown in \cite{FS-FBMC2012}, a better channel equalization can be achieved through frequency spreading implementation of FBMC systems compared with the FBMC implementations based on polyphase networks. Hence, based on the results of \cite{FS-FBMC2012} and \cite{FBMCMassive2014}, we are interested in the investigation of the FS-FBMC in massive MIMO channels. As will be shown in Section~\ref{sec:NR}, due to the higher frequency resolution of FS-FBMC, the subcarrier bands in these systems can be widened even more than what is reported in \cite{FBMCMassive2014}. Thus, the latency due to the synthesis and analysis filter banks can be reduced to a great extent compared with single antenna systems.

In this section, we derive the mathematical foundation for MMSE-FSE in massive MIMO systems. In a multiuser MIMO system similar to the one in \cite{Marzetta2010}, let $N_{\rm r}$ be the number of receive antennas at the base station and $M$ the number of users. Each user's mobile terminal (MT) is equipped with a single transmit/receive antenna. All MTs are simultaneously communicating with the BS in \textit{time division duplexing} (TDD) mode. The number of BS antennas is considered much larger than the number of users, i.e., $N_{\rm r}\gg M$. The users communicate with the BS using CMT modulation. The users' signals are distinguished by their respective channel gains between each MT antenna and the BS antennas. The channel gain vectors for different users are assumed to be independent with respect to each other. Recalling the structure of the CMT receiver from Section~\ref{sec:FSCMT}, the output at the $i^{\rm th}$ bin of the FFT block from the $\ell^{\rm th}$ user at all the receive antennas can be stacked into an $N_{\rm r}\times 1$ vector $\tilde{\br}_i^\ell = r_i^\ell\bh_i(\ell)$ where $\bh_i(\ell)$ is the channel gain vector whose elements are the gains between the $\ell^{\rm th}$ MT antenna and the BS antennas at the $i^{\rm th}$ frequency bin. The coefficient $r_i^\ell$ is the received signal at the $i^{\rm th}$ bin of the output of the FFT block at the receiver from user $\ell$ in an ideal channel. The vectors $\tilde{\br}_i^\ell$ are the received signals from different users that contribute to form the received signal at the BS. Accordingly, the received signal vector can be written as   
\be\label{eqn:ri}
\tilde{\br}_i=\sum_{\ell=0}^{M-1}{\tilde{\br}_i^\ell}+\bv_i=\bH_i\br_i+\bv_i,
\ee
where $\bH_i=[\bh_i(0),\ldots,\bh_i({M-1})]$, $\br_i=[r_i^0,\ldots,r_i^{M-1}]^{\rm T}$ and $\bv_i$ is an $N_{\rm r}\times 1$ vector that contains the FFT samples of the additive white Gaussian noise (AWGN) at the $i^{\rm th}$ bin of the output of the FFT blocks at different receive antennas. 

In order to estimate different users' transmitted symbols, a number of equalization techniques can be utilized; namely, matched filter (MF) and MMSE. However, as it is noted in \cite{FBMCMassive2014}, in realistic situations when the number of BS antennas is finite, MF equalization incurs some performance degradation due to the residual multiuser interference. In this paper, our discussion will be limited to the MMSE-FSE which is known to be a superior technique. 

In MMSE-FSE for massive MIMO systems, similar to single antenna FS-FBMC systems, equalization needs to be performed before despreading the output of the FFT blocks. Therefore, the MMSE linear combining aims at estimating the vector $\br_i$ that contains the equalized samples of different users at the output of their FFT blocks. The MMSE solution of (\ref{eqn:ri}) is optimal as it maximizes the signal-to-interference-plus-noise (SINR),\cite{FarhangBook2013}. Following the same line of derivations as in \cite{FBMCMassive2014}, the MMSE estimates of $\br_i$'s can be obtained as
\be \label{eqn:rhat}
\hat{\br}_i=\bW_i^{\rm H}\tilde{\br}_i,
\ee
where the coefficient matrix $\bW_i=\bH_i(\bH_i^{\rm H}\bH_i+{\sigma_v^2}\I_{M})^{-1}$ contains the optimal MMSE filter tap weights for different users in its columns. It is worth mentioning that the elements of the noise vector $\bv_i$ are assumed to be independent and identically distributed Gaussian random variables with variances of $\sigma_v^2$ and $\mathbb{E}\left[\bv_i\bv_i^{\rm H}\right]=\sigma_v^2\I_M$. After the MMSE estimation of the vectors $\hat{\br}_i$ for $i=0,\ldots,N-1$, frequency despreading can be separately applied for each user. To this end, let $\hat{\bR}=[\hat{\br}_0,\ldots,\hat{\br}_{N-1}]^{\rm T}$. Consequently, the MMSE estimates of the transmitted data symbols of different users, i.e., $\bs_\ell$ for $\ell=0,\ldots,M-1$ can be obtained through frequency despreading
\be \label{eqn:shat}
\hat{\bS}=\Re\{\bPhi^{-1}\bA^{\rm T}\hat{\bR}\},
\ee
where the $L\times M$ matrix $\hat{\bS}$ contains the MMSE estimation of the transmitted data symbols of different users, i.e., $\hat{\bs}_\ell$'s, on its columns.

Based on equations (\ref{eqn:ri}) to (\ref{eqn:shat}) and the assumption of having a flat fading channel gain over each frequency bin of DFT, the output SINR of the CMT receiver with MMSE channel equalization for the $\ell^{\rm th}$ user at subcarrier $i$ can be calculated as follows.
\be\label{eqn:SINR_MMSE}
{\rm SINR}_{\rm MMSE}^{(\ell,i)}=\frac{P_{\rm s}^{(\ell,i)}}{P_{\rm I}^{(\ell,i)}},
\ee
where $P_{\rm s}^{(\ell,i)}$ and $P_{\rm I}^{(\ell,i)}$ are signal and interference-plus-noise powers, respectively. $P_{\rm s}^{(\ell,i)}$ and $P_{\rm I}^{(\ell,i)}$ can be obtained using equations (\ref{eqn:PS}) and (\ref{eqn:PI}) that are shown on the top of the previous page where $\bw_{i,{\Re}}(m)$ and $\bh_{i,{\Re}}(m)$ represent the real parts of the $m^{\rm th}$ columns of the matrices $\bW_i$ and $\bH_i$, respectively. Similarly, the subscript $\Im$ in the equations (\ref{eqn:PS}) and (\ref{eqn:PI}) shows the imaginary part of the corresponding vectors. It is worth mentioning that equation (\ref{eqn:SINR_MMSE}) is used as a benchmark to investigate the channel flatness assumption in Section~\ref{sec:NR}.

\section{Numerical Results}\label{sec:NR}

\begin{figure}
\centering 
\subfigure[]{\hspace{-7 mm}
\includegraphics[scale=0.6]{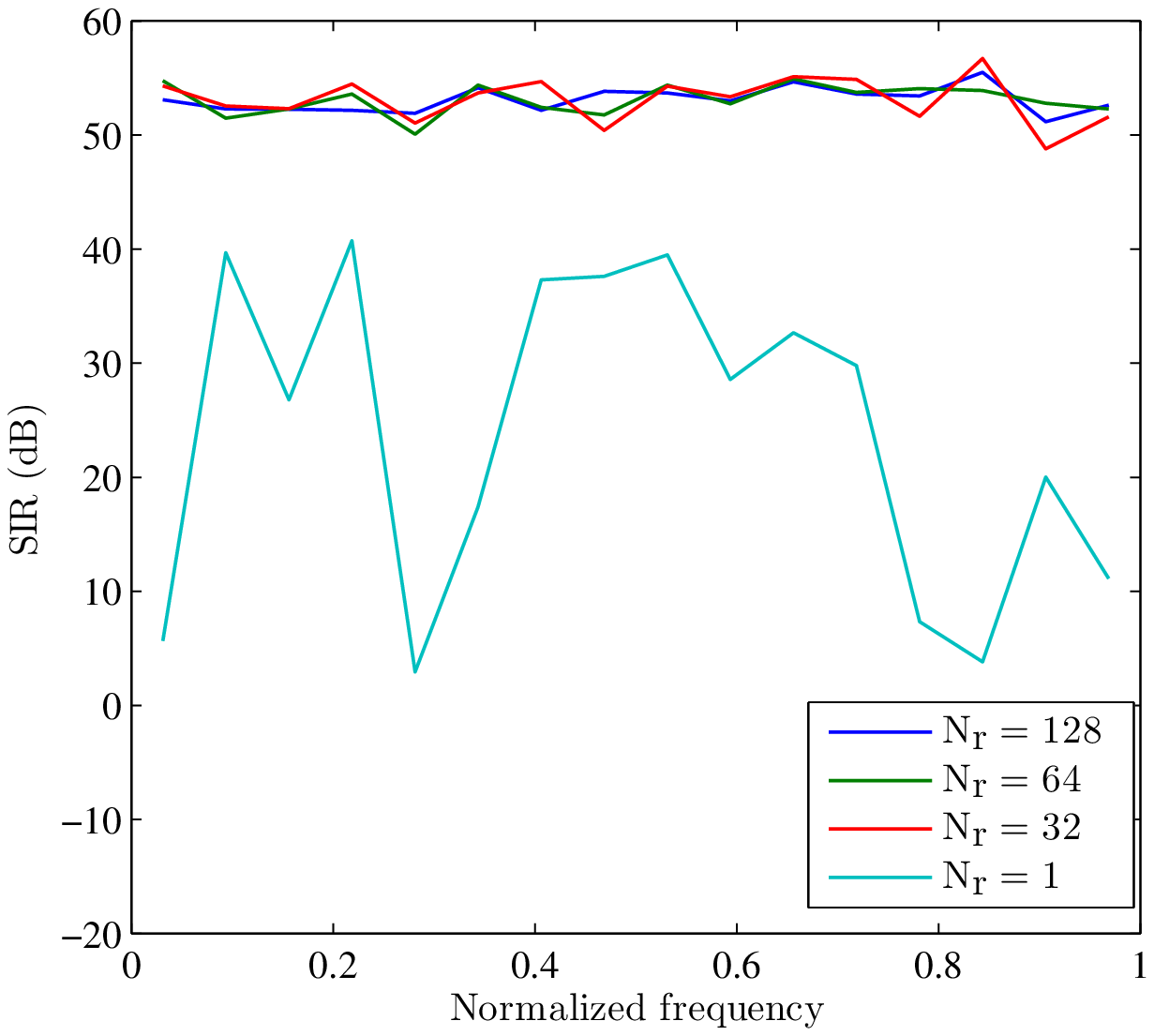}
\label{fig:L16}
}
\subfigure[]{\hspace{-7 mm}
\includegraphics[scale=0.6]{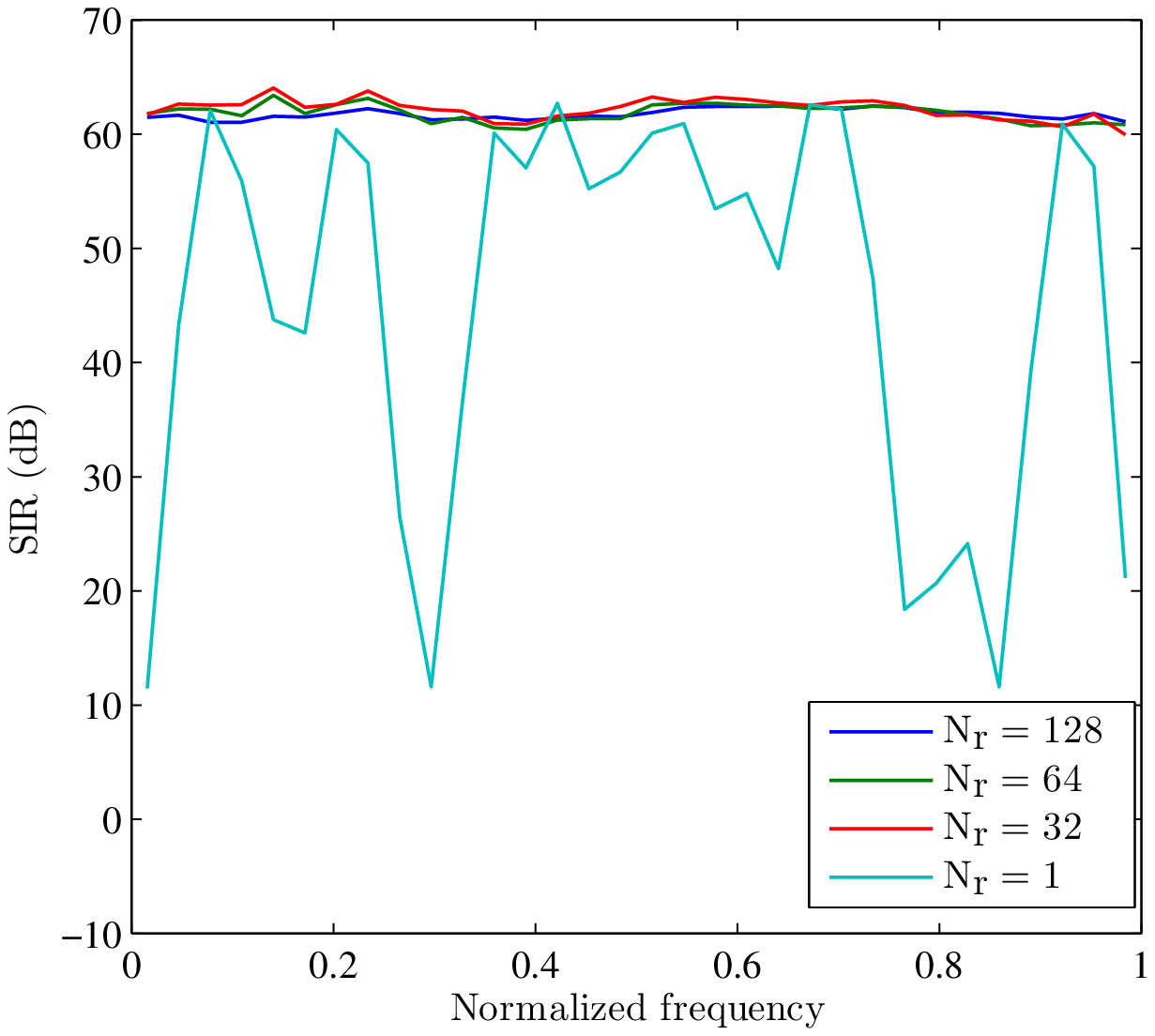}
\label{fig:L32}
}
\subfigure[]{\hspace{-7 mm}
\includegraphics[scale=0.6]{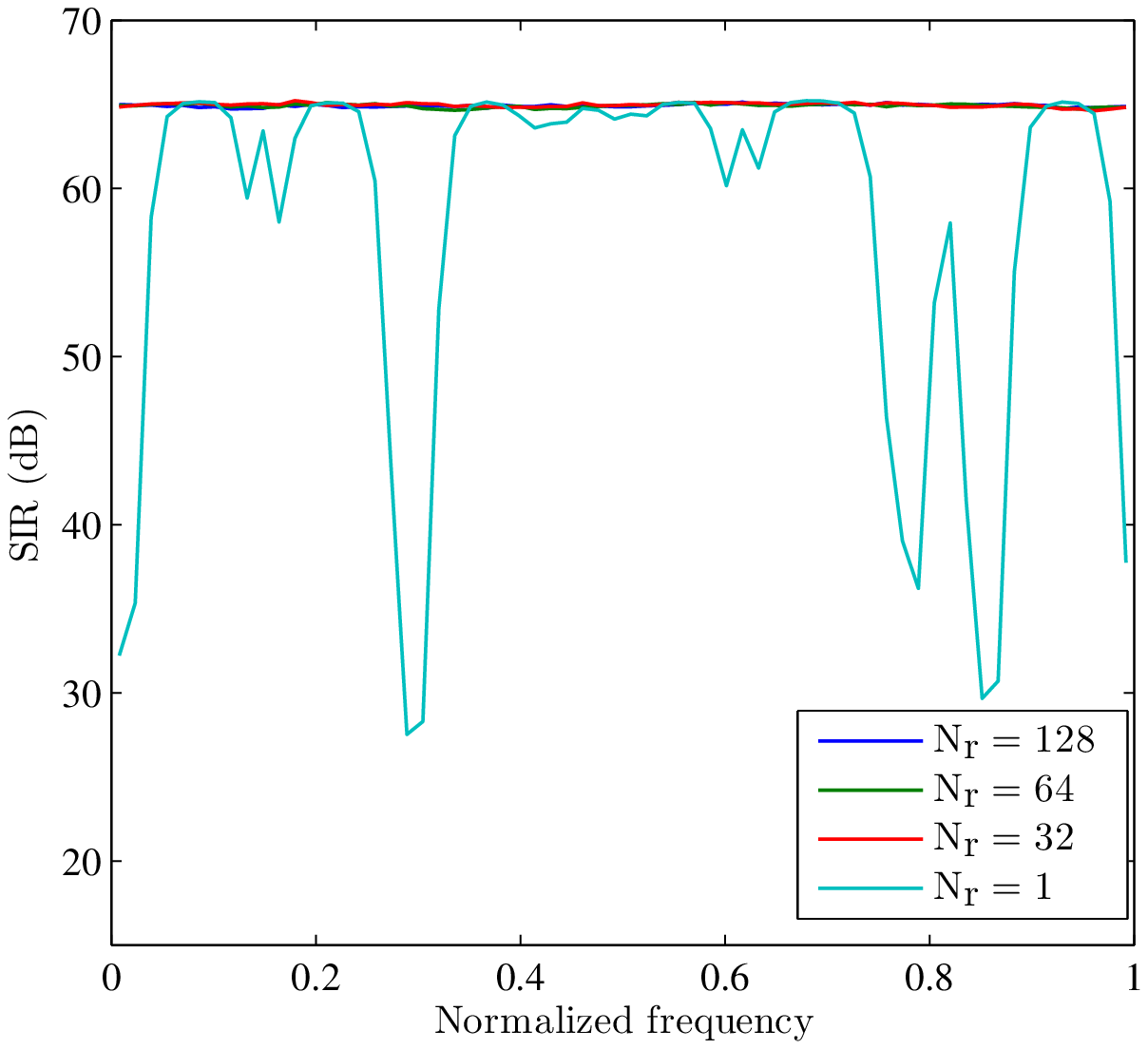}
\label{fig:L64}
}
\caption{\subref{fig:L16}, \subref{fig:L32} and \subref{fig:L64} compare the SIR performance for the cases of $8,16$ and $32$ subcarriers, respectively, with different number of receive antennas, $N_{\rm r}$. }
\label{fig:SIR}
\end{figure}

In this section, the theoretical developments of the paper are analyzed and corroborated through numerical results. We use the results of \cite{FBMCMassive2014} as the basis in order to evaluate the signal processing power of the FS method in the context of massive MIMO. It is worth noting that all of our simulations are based on a sample set of channel responses generated according to the SUI-4 channel model proposed by the IEEE802.16 broadband wireless access working group, \cite{SUI}. Additionally, the channels between different users and different antennas are considered independent with respect to each other.

In the first set of simulations (Fig.~\ref{fig:SIR}), we study the effect of increasing the number of antennas on the signal-to-interference ratio (SIR) performance. In Fig.~\ref{fig:SIR}, a single-user in a noise-free channel is considered to investigate the self-equalization property in FS-FBMC based massive MIMO systems. To evaluate the robustness of the FSE to the subcarrier width, we have repeated the simulations for three cases of 8, 16, and 32 subcarriers and different number of receive antennas at the base station. More specifically, for the total bandwidth of 2.8 MHz, we consider the subcarrier widths of 350, 175, and 87.5 kHz, respectively. SIRs are evaluated at all the subcarrier channels. Note that in each curve, the number of points along the normalized frequency axis is equal to the number of subcarriers, $L$. As these results highlight, we can achieve a significant performance improvement even for the subcarrier spacings as large as 350 kHz. 

\begin{figure}[!tb]
\centering
\includegraphics[scale=0.60]{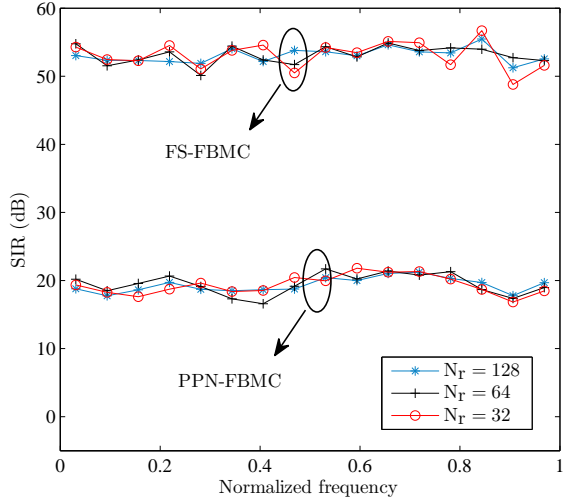}
\vspace{0.03in}
\caption{SIR  for the case of having $L=16$, $K=4$, $M=1$ and different number of BS antennas.}
\label{fig:SIR_SingleUser}
\end{figure}

The effectiveness of these results will be more clear if we compare the FSE technique with the conventional PPN algorithm. Fig.~\ref{fig:SIR_SingleUser} compares the SIR performance of the PPN-FBMC with FS-FBMC system. A single-tap equalizer per subcarrier is used in the PPN-FBMC structure. From Fig.~\ref{fig:SIR_SingleUser}, it can be understood that the SIR improvement of higher than $30$ dB can be achieved through FS-FBMC in comparison with the PPN-FBMC structure. This means that higher order modulation schemes can be utilized in FS-FBMC based massive MIMO systems compared with their PPN-based counterpart. The same as the previous case, the total bandwidth is fixed to $2.8$ MHz, the number of subcarriers $L=16$ and overlapping factor of $K=4$ are considered. This results in the subcarrier spacing of $2800/16=175$ kHz. This subcarrier spacing is relatively wide. In fact, it is $175/15 \approx 12$ times larger than the subcarrier spacing of the OFDM-based systems (e.g. IEEE 802.16 and LTE). 

\begin{figure}[!tb]
\centering \vspace{0.15in}
\includegraphics[scale=0.6]{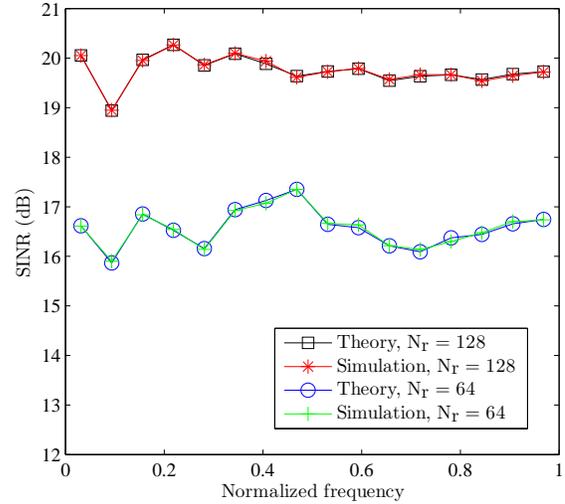} 
\caption{SINR evaluation of the MMSE linear combining for the case of $L=16$, $K=4$, and $M=6$. The SNR at the receiver input is $-1$ dB.}
\label{fig:SINR_MultiUser}
\end{figure}

Next, we consider a multiuser scenario with $M=6$ users, $L=16$ subcarriers and $K=4$. With the assumption of having perfect power control for all the users, the signal-to-noise-ratio (SNR) of $\textrm{SNR}_{\rm in}=-1$ dB at the input of BS antennas is considered. The output SINR can be obtained using $\textrm{SNR}_{\rm in} + 10 \log_{10} N_{\rm r}$ in dB. Fig.~\ref{fig:SINR_MultiUser} compares the SINR values from simulation for two cases of $N_{\rm r}=128$ and $N_{\rm r}=64$ with the theoretical SINR values that are calculated based on the assumption of having a flat fading channel per subcarrier band that is calculated from equation (\ref{eqn:SINR_MMSE}). As the figure depicts, the theoretical derivations match perfectly with the simulation results. This emphasizes that the channel is flattened over all the subcarrier bands, thanks to the self-equalization property. The subcarrier spacing that achieves the same SINR as that of (\ref{eqn:SINR_MMSE}) is $4$ times larger than what was suggested in \cite{FBMCMassive2014} which brings a significant improvement in terms of shortening the length of the prototype filter and hence reducing the latency that is imposed by the transmit and receive filter banks.

\section{Conclusion}\label{sec:Conclusion}
In this paper, the idea of frequency spreading FBMC was extended to the massive MIMO systems. In addition, an effective MMSE equalization scheme for FBMC-based massive MIMO systems was derived. It was shown that further subcarrier widening than what was suggested in \cite{FBMCMassive2014} is possible using FS-FBMC structure while gaining from self-equalization property. Based on the SIR and SINR performance results, higher order modulation techniques can be utilized in FS-FBMC based massive MIMO systems. Furthermore, improvements in terms of bandwidth efficiency, robustness to frequency offsets, complexity, PAPR and latency compared with the polyphase based FBMC systems can be achieved using FS-FBMC structure.

\bibliographystyle{IEEEtran}

\end{document}